\documentclass[a4paper,11pt]{article}
\usepackage{pos}
\usepackage{todonotes}
\usepackage{siunitx}
\usepackage{hyperref}
\usepackage{wrapfig}
\usepackage{lineno}
\usepackage{comment}

\title{Reconstruction of Neutrino Events in IceCube using Graph Neural Networks
}
 \ShortTitle{Graph Neural Network reconstruction in IceCube}

\author{The IceCube Collaboration \\{\normalsize \normalfont(a complete list of authors can be found at the end of the proceedings)}}

\emailAdd{martin.haminh@icecube.wisc.edu}

\abstract{
The IceCube Neutrino Observatory is a cubic-kilometer scale neutrino detector embedded in the Antarctic ice of the South Pole. In the near future, the detector will be augmented by extensions, such as the IceCube Upgrade and the planned Gen2 detector. The sparseness of observed light in the detector for low-energy events, and the irregular detector geometry, have always been a challenge to the reconstruction of the detected neutrinos' parameters of interest. This challenge remains with the IceCube Upgrade, currently under construction, which introduces seven new detector strings with novel detector modules. The Upgrade modules will increase the detection rate of low-energy events and allow us to further constrain neutrino oscillation physics. However, the geometry of these modules render existing traditional reconstruction algorithms more difficult to use. We introduce a new reconstruction algorithm based on Graph Neural Networks, which we use to reconstruct neutrino events at much faster processing times than the traditional algorithms, while providing comparable resolution. We show that our algorithm is applicable not only to reconstructing data of the current IceCube detector, but also simulated events for next-generation extensions, such as the IceCube Upgrade.

\vspace{4mm}
{\bfseries Corresponding authors:}
Martin Ha Minh$^{1*}$\\
{$^{1}$ \itshape Technical University Munich}\\
$^*$ Presenter

\FullConference{37$^{\rm{th}}$ International Cosmic Ray Conference (ICRC 2021)\\
		July 12th -- 23rd, 2021\\
		Online -- Berlin, Germany}

}

\begin{document}
\maketitle

\section{Introduction}
\subsection{The IceCube detector and IceCube Upgrade}
The IceCube Neutrino Observatory \cite{icecube} is a cubic-kilometer scale neutrino detector embedded in the Antarctic ice of the South Pole. Neutrinos interact inside or close to the detector volume and produce secondary particles that can emit Cherenkov light. A fraction of this light is then collected by the 
photo multipliers (PMTs) \cite{pmts} in our detector modules, also known as Digital Optical Modules (DOMs), which are situated on 86 detector strings, with 5160 DOMs in total. One region within the IceCube detector is denser both in the horizontal spacing of the strings and the vertical spacing between DOMs, compared to the rest of the detector volume. The higher density of photo-cathode area pushes the energy threshold for detected neutrinos below the first atmospheric neutrino oscillation maximum at \SI{25}{\giga\electronvolt}, down to \SI{10}{\giga\electronvolt}. This region is called DeepCore \cite{deepcore} and is used for oscillation  \cite{osc_prl} \cite{nutau} and beyond standard model physics \cite{sterile}. 

\begin{wrapfigure}{r}{0.5\textwidth}
    \centering
    \includegraphics[width=.5\textwidth]{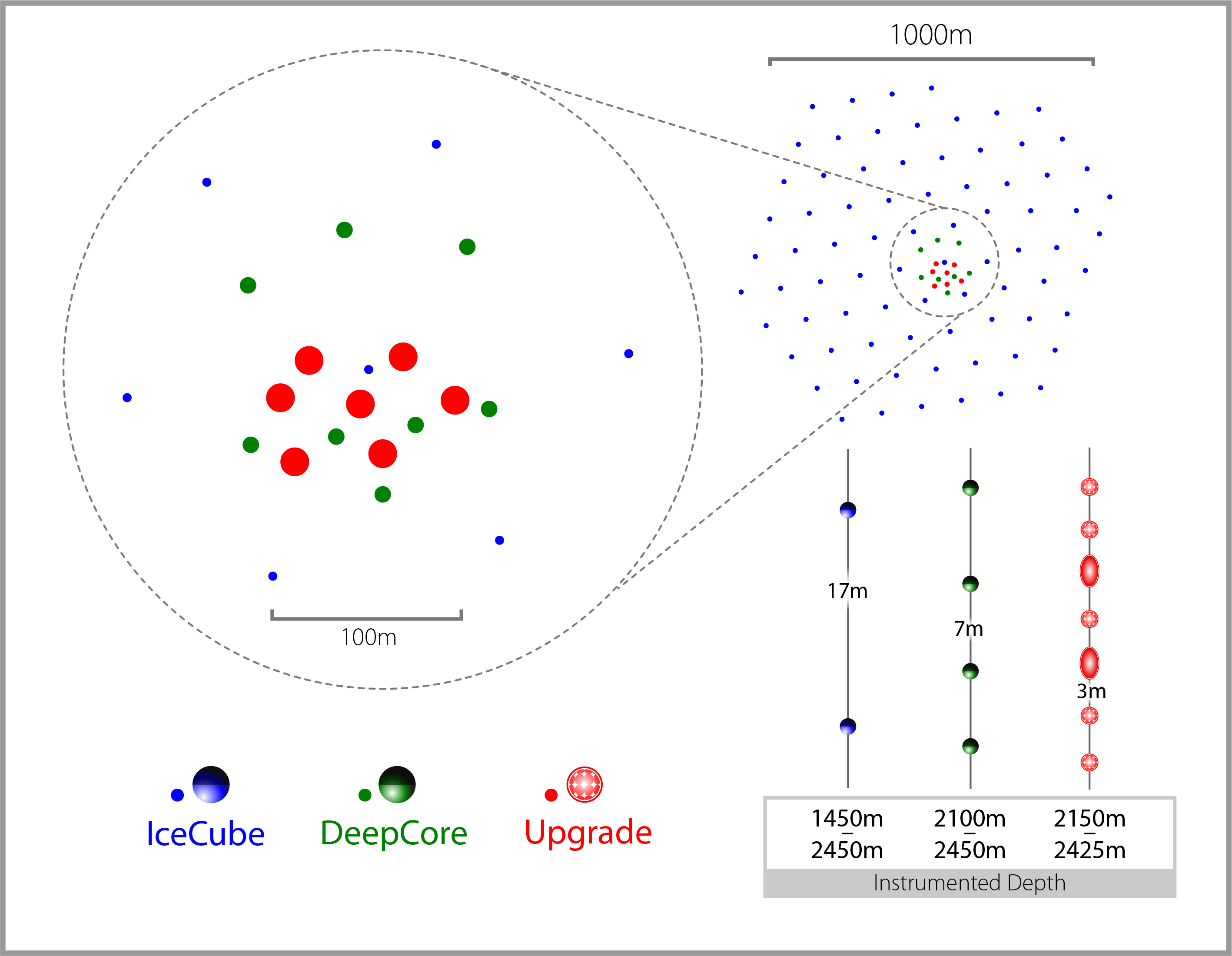}
    \caption{Layout for the IceCube, DeepCore, and planned Upgrade strings.}
    \label{fig:stringlayout}
    \vspace{-2ex}
\end{wrapfigure}

An additional low-energy extension to the detector, called the IceCube Upgrade \cite{Ishihara:2019aao}, is currently in construction and will improve our sensitivity in oscillation physics. This extension comprises 7 strings even denser in spacing than the DeepCore strings. The strings will be outfitted with calibration devices, as well as novel detector modules. While the current DOMs exclusively contain downward-pointing PMTs, the new models will also feature PMTs that cover an almost $\mathrm{4\pi}$ solid angle region to improve the directional resolution of arriving neutrinos. The IceCube Upgrade is planned to go into operation in 2023/2024.

\subsection{Neutrino oscillation physics}

Neutrinos produced in the earth's atmosphere can travel through the entire earth's volume. While traveling, they oscillate between the different flavors: electron ($\nu_e$), muon ($\nu_\mu$), and tau ($\nu_\tau$). Under standard assumptions about the oscillation parameters of the neutrino, the density profile of the earth, and the atmospheric particle flux model, we can make a prediction about the appearance rate of a neutrino with a set of given experimental observables inside the IceCube detector. These properties are the neutrino's energy, traveled distance, flavor, and whether it underwent a charged current (CC) or neutral current (NC) interaction.

As we assume symmetry in the azimuthal direction, the travel distance can also be expressed by the zenith angle of the neutrino arrival direction in the detector. Additionally, we do not have a direct handle on the flavor and the interaction type of a neutrino event. However we primarily encounter two different event topologies: track(-like) and cascade(-like) events. Charged-current $\nu_\mu$ interactions, as well as a fraction of $\nu_\tau$ interactions, will imprint themselves as track events. All other flavors and interaction types will appear as cascade events. In summary, to perform an oscillation analysis in IceCube, we have to estimate the neutrino energy, the zenith angle of the arrival direction, and the event topology of each neutrino interaction.

\section{Low-energy neutrino event reconstruction}

\subsection{Current methods}
The current baseline approach uses a likelihood-based method using "photon tables". 
Using a parametrization that describes a neutrino event, we can predict the number of Cherenkov photons and the direction they will be emitted in. The photon tables then describe the propagation of those photons through the detector. To reconstruct an event, the algorithm then minimizes the likelihood between said recorded event and photon table predictions as a function of event parameters. While this method works, there are certain issues. Most of these can be summed up with computational cost.

\begin{wrapfigure}{l}{0.5\textwidth}
    \centering
    \includegraphics[width=.45\textwidth]{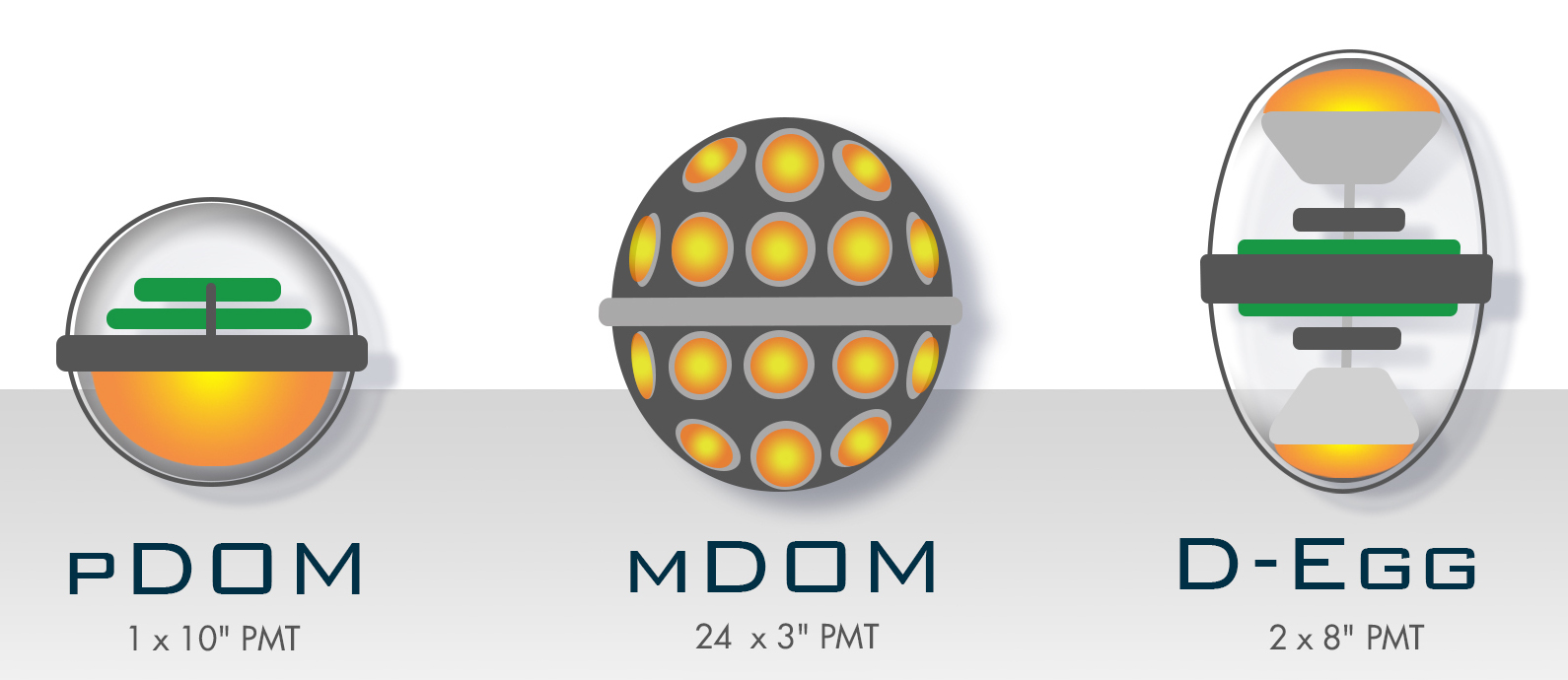}
    \caption{New DOM architectures for the Upgrade strings.}
    \label{fig:newdoms}
\end{wrapfigure}

The reconstruction of one event using the baseline method takes approximately \SI{40}{\second}. As we handle numbers of events on the order of $\mathrm{10^7}$, the time necessary compounds very quickly; calling for weeks of HPC cluster usage. Additionally, the current photon tables assume only downwards-facing PMTs.
We would therefore have to expand the already large photon tables by another dimension to accommodate for the additional PMT directions. The current photon tables require compression, after which they are still a few GB in size. While they currently fit into the memory of most systems, this cannot be expected with the additional dimension. The expansion of photon tables will also increase the time needed for the minimization of each event reconstruction likelihood, exacerbating the first point. The need for more efficient algorithms apparent.

An additional benefit of a faster reconstruction algorithm is that it can be used earlier in the data processing chain. Currently, our events are filtered to remove background events due to muons and noise before they are passed to the reconstruction algorithm. This is due to the aforementioned time it takes to reconstruct a large amount of events. With a faster algorithm, we can reconstruct the events earlier in the processing chain and could in addition use the reconstructed parameters to improve the data selection. 

Machine learning approaches excel with respect to speed and flexibility. These approaches, specifically deep neural networks, have found many applications in the current physics landscape, also in the IceCube collaboration \cite{dnnpaper}\cite{maxi}. Popular machine learning approaches include Convolutional Neural Networks (CNNs) or Multilayer Perceptrons (MLPs). These however have caveats, such as the requirement of fixed-size inputs, or the assumption of symmetries in our detector.

\subsection{Graph Neural Networks}

Graph Neural Networks (GNNs) are a good candidate to amend these issues. 
In GNNs the information is represented by a collection of nodes These nodes can then be connected by edges, based on their relationship to each other. 
The convolutional layer then collects the information of the neighbors around each node, as described by the edges. This approach allows us to encode information in an irregular configuration, as opposed to CNNs, that are limited by their regular convolutional kernel.

GNNs have been used previously in IceCube, but only for signal classification \cite{nersc}. The approach presented in this proceeding is among the first to employ GNNs for reconstruction and interaction type classification.


\section{Description of technique}


\begin{wrapfigure}{r}{.5\textwidth}
    \vspace{-2ex}
    \centering
    \includegraphics[width=.5\textwidth]{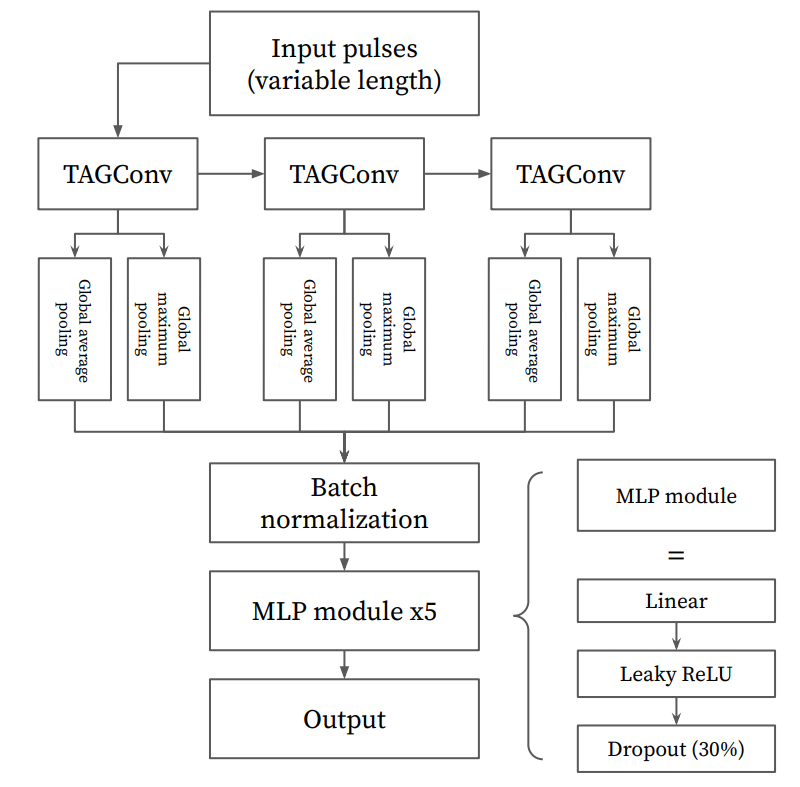}
    \caption{Schematic of the network architecture used.}
    \label{fig:architecture}
\end{wrapfigure}

\subsection{Building the graphs}
We propose a method that works on a per-pulse basis. Each detector event is described a as a pulse series, i.e.\ a list of pulses. In the simplest way, each pulse can be described by following quantities: the hit DOM, and therefore the position of the pulse, collected charge, and time recorded. For the IceCube Upgrade, we also introduce the pointing direction of the PMT, as well as the type of optical module that registered the charge. We can then represent each event by a graph, with the nodes representing the pulses in an abstract 5-dimensional (or 12-dimensional for the Upgrade) space. The next task is to connect these nodes to each other with edges. We use the $k$ nearest neighbors calculated from the Euclidean distance in position and time to decide which nodes should be connected by edges. The edges are bidirectional and unweighted.

\subsection{Graph convolutional layers}
After building the graphs, we can apply various convolutional layers. In essence, convolutional layers extract the information about a given input and expand them onto additional dimensions. In our first efforts, we found the 'topology adaptive graph convolutional network' layer (TAGCN) \cite{tagconv} in the \texttt{PyTorch Geometric} package \cite{Fey2019FastGR} to fit our needs the best. TAGCN provides a fixed-size, CNN-like filter, expanding established methods to the graph language. In our approach, we employ a combination of average pooling and maximum pooling.

\subsection{Graph pooling and decoding}
Traditional neural network methods, such as CNNs or MLPs, usually require a fixed input size. This makes per-pulse approaches difficult, as the number of pulses per event varies. GNNs however, provide a method called graph pooling. By pooling a graph, the graph information is shaped to fit a given input size, by e.g.\ taking the average, the maximum, or the sum of a set of values. We perform this step after applying the graph convolutional layers. This way we circumvent losing information by prematurely summarizing our input. The output is then passed to MLPs for additional learning and decoding to our parameters of interest.

\subsection{Training targets and loss functions}
In our oscillation analysis, we are primarily interested in three quantities: neutrino energy, the zenith angle of the arrival direction, and the event topology. The energy range we operate in spans from approximately 1 GeV to 1000 GeV. To provide a loss that is not biased towards higher energies, we train on the logarithm of the energy in base 10 ($\log_{10}(E)$), in combination with a mean squared error (MSE) as the loss function. For the zenith angle, we found the conversion of the angle $\theta$ to a 2D cartesian $[\sin(\theta), \cos(\theta)]$ and then using an MSE loss function to yield the best results. For the event topology, we want to distinguish between two signatures, track and cascade event hypothesis, so we choose a binary cross entropy loss.

A schematic for the network is shown in Fig.\ \ref{fig:architecture}.

\section{Results}
We split the application of our method into two parts: The first part details the application of our reconstruction to simulation of our current oscillation analysis, as well as a comparison to our baseline reconstruction algorithm. In the second part we apply the GNN algorithm to the simulation of the upcoming IceCube Upgrade and illustrate the improvement in resolution coming from the additional strings and detector modules.

\subsection{Application to the current IceCube oscillation analysis}

\begin{figure}[hbt!]
    \vspace{-2ex}
    \centering
    \includegraphics[width=.9\textwidth]{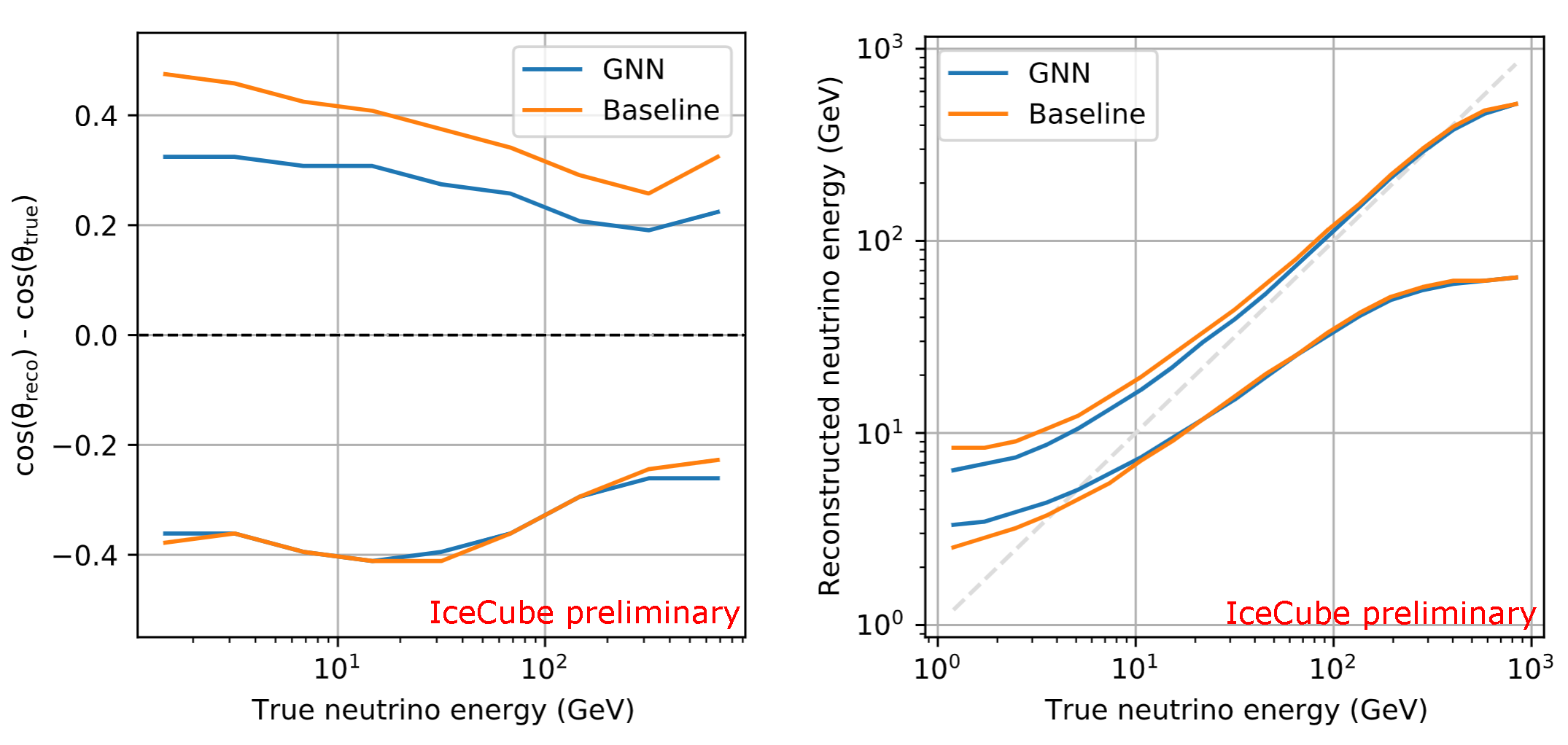}
    \caption{Bands comparing the resolution vs energy of our GNN with the baseline algorithm, for the zenith angle $\cos(\theta)$ (left) and the neutrino energy $E$ (right). 68\% of the best reconstructed events lie between the plotted bands. The dashed lines indicate the target values.}
    \label{fig:oscnext_resolution}
\end{figure}

For training, we use a mixture of Monte Carlo simulations of $\nu_e$, $\nu_\mu$, and $\nu_\tau$ interaction events. The simulation events have undergone a removal of noise pulses using an algorithm. For testing we use the final event selection also used for our oscillation analyses, which includes the removal of background events from muons and noise. We then predict our parameters of interest, energy $E$, zenith angle $\theta$, and event topology, and compare them to the Monte Carlo truths. We can also compare our results to the resolution of our baseline algorithm based on photon tables for reconstructing the energy and the zenith angle. Results are shown in Fig.\ \ref{fig:oscnext_resolution}.

For both resolution in zenith angle and energy, we can see that the 68\% bands of GNN-reconstructed events for are generally more constraining than the bands for the baseline, especially at lower neutrino energies. This result is especially satisfactory, as it is not only an overall improvement, but also specifically for neutrino events with energies that are usually harder to reconstruct, due to the low number of recorded pulses. 

The distorted shape of the bands for the reconstructed energy stems from two sources. In the energy range below \SI{10}{\giga\electronvolt} our selection is biased towards higher reconstructed values due to the hard lower threshold at 8 pulses. On the other hand, above \SI{100}{\giga\electronvolt}, a fraction of the neutrino energy is transferred into secondary particles that do not emit Cherenkov light and are therefore invisible to the detector.

Over all tested samples, the standard deviation of the difference between true and reconstructed $\cos(\theta)$ for the GNN is $\sigma_{\mathrm{\cos(\theta),\ GNN}}= 0.36$ and for the baseline algorithm $\sigma_{\mathrm{\cos(\theta),\ Retro}}= 0.44$. Due to the aforementioned distortion in the energy bands, we select the true neutrino energy range between 10 and \SI{100}{\giga\electronvolt} for the evaluation of the energy resolution. There, for the standard deviation of the difference between true and reconstructed $\mathrm{log_{10}}$(energy) we obtain $\sigma_{\mathrm{log_{10}(E),\ GNN}}= 0.24$ for the GNN and $\sigma_{\mathrm{log_{10}(E),\ Retro}}= 0.26$ for the baseline algorithm.

\begin{wrapfigure}{r}{0.5\textwidth}
    \vspace{-6ex}
    \centering
    \includegraphics[width=.45\textwidth]{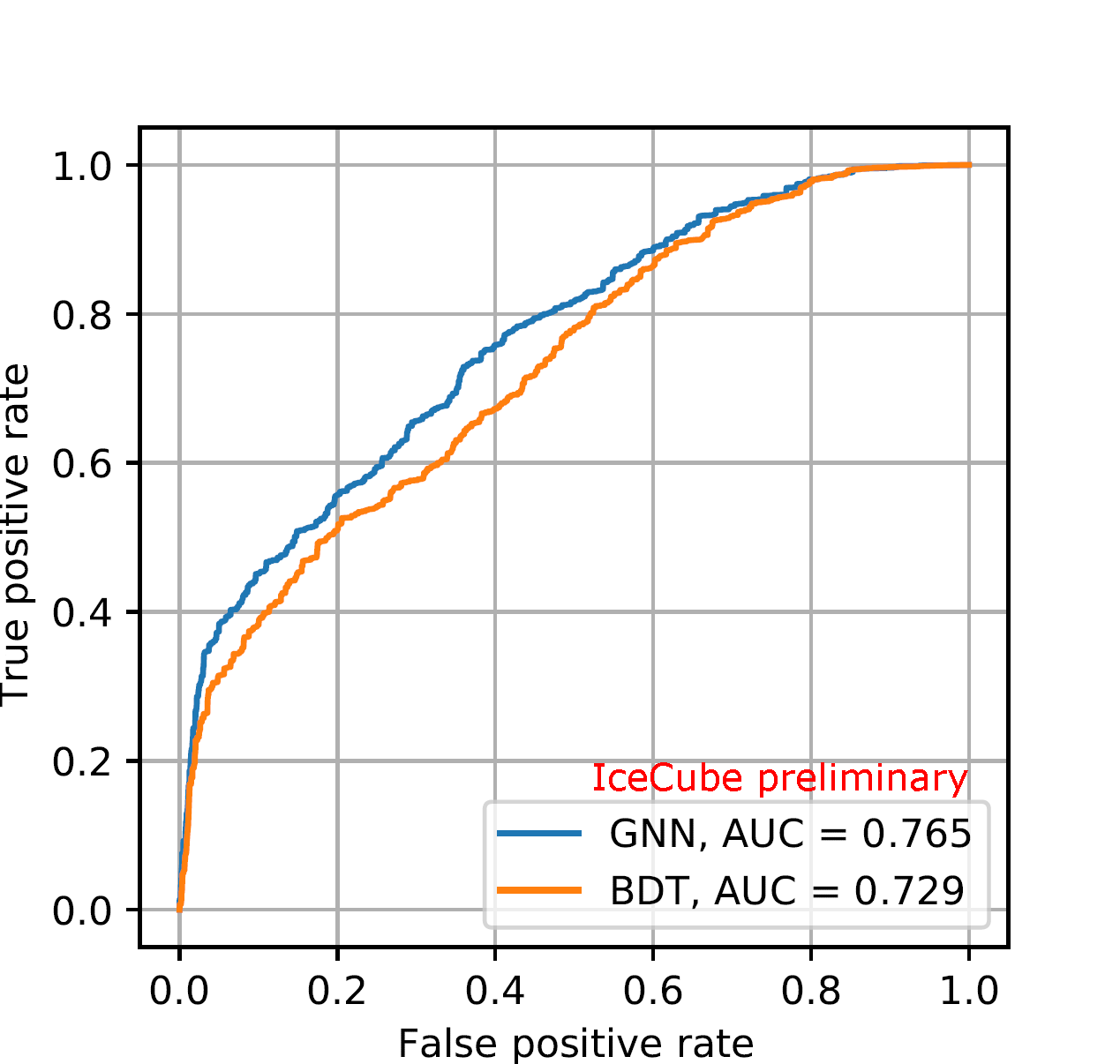}
    \caption{Receiver operating characteristic curve with area under curve (AUC) for a track--cascade identifier using the GNN and the baseline algorithm (BDT).}
    \label{fig:oscnext_pid}
    \vspace{-2ex}
\end{wrapfigure}

We then compare the prediction of the event topology with the baseline algorithm based on a Boosted Decision Tree (BDT). \mbox{Fig.\ \ref{fig:oscnext_pid}} shows the receiver operating characteristic curve for the event topology classifiers using the GNN and the BDT respectively. We can see that the ratio between true positive rate and false positive rate for all thresholds is improved with the GNN, leading to a higher area under the curve (AUC).

Another improvement is the speedup compared to the current algorithms. With our GNN, it is possible to reconstruct one event within less than \SI{3}{\milli\second}, offering a speedup on the order of $10^4$. This leads to a reconstruction time of circa \SI{6}{\hour} for the whole simulation dataset on only one GPU. The same reconstruction takes approximately two weeks using the full IceCube HPC cluster.

In summary, our GNN reconstruction offers not only improvement in the resolution of the zenith angle of the direction, the energy and the event topology of the neutrino event, but also a decrease in reconstruction time.

\subsection{Application to the future IceCube Upgrade}

\begin{wrapfigure}{l}{0.45\textwidth}
    \centering
    \vspace{-10pt}
    \includegraphics[width=.45\textwidth]{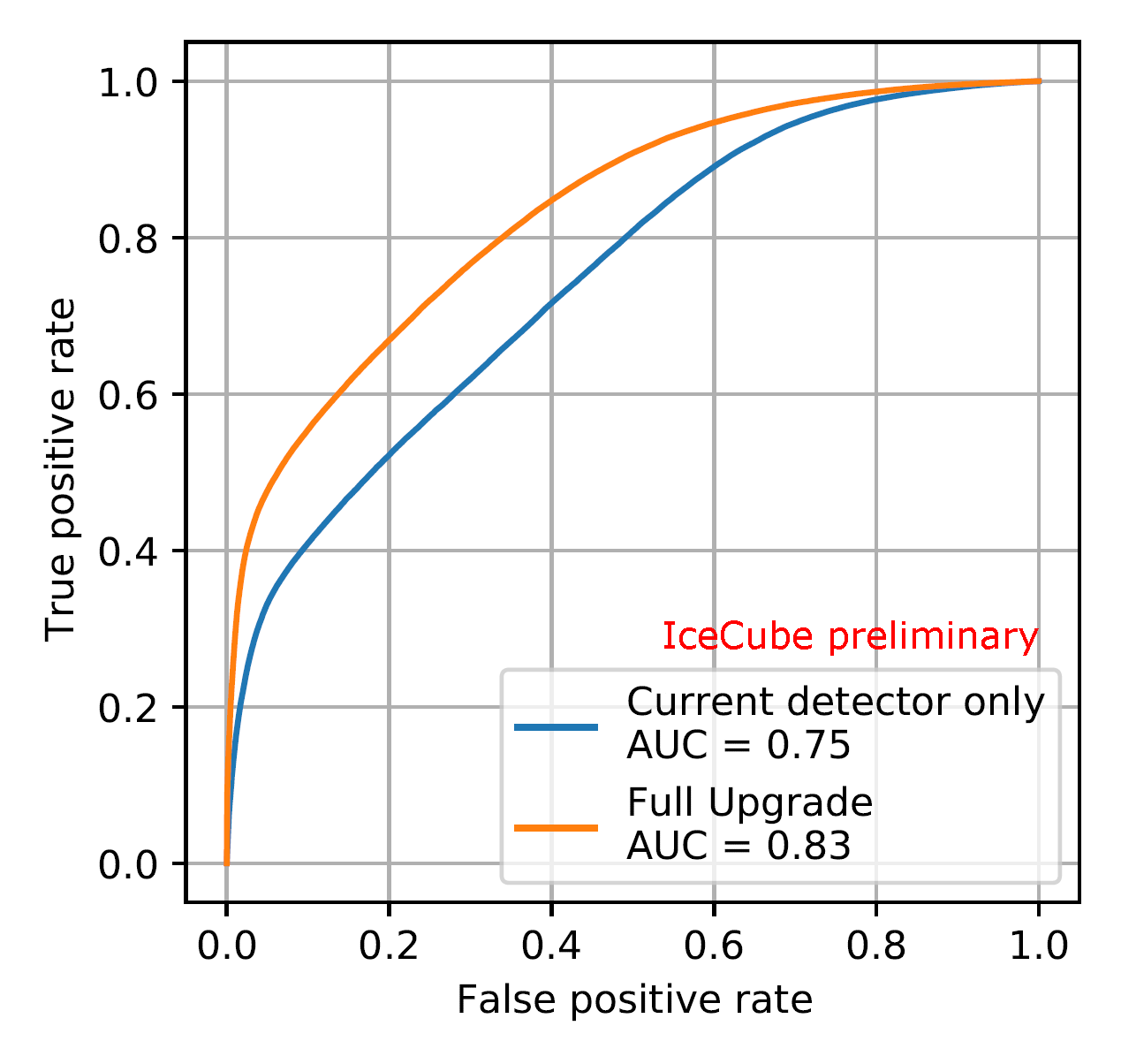}
    \vspace{-20pt}
    \caption{Receiver-operator-characteristic curve and area under curve (AUC) for a track--cascade identifier using the full Upgrade detector and the current detector configuration.}
    \label{fig:upgrade_roc}
    \vspace{-2.5ex}
\end{wrapfigure}

To show the flexibility of our approach, we now use the GNN to reconstruct events simulated for our future detector upgrade. As this extension also includes new detector models, we adjust the input parameters to include additional information: There are three new detector models, namely the pDOM, the mDOM, and the D-Egg. The mDOM and the D-Egg include PMTs that not only point downwards, but in multiple other directions. So besides the previously used $x$-, $y$-, and $z$-position of the recorded pulse, as well as its collected charge and time of recording, we also include the detector module type and the direction of the PMT that recorded the pulse.

We train the networks on a dataset, algorithmically cleaned of noise pulses, but without major background rejection. The dataset to test the networks are based on the same processing as the training set, as the event selection for IceCube Upgrade is still in its early development phase. We are also interested to see the improvement coming from the Upgrade. To do so, we remove pulses recorded by the new modules, and reconstruct those events with the networks trained in the previous section. This way we have a direct event-by-event comparison.

\begin{figure}[hbt!]
    \centering
    \includegraphics[width=.95\textwidth]{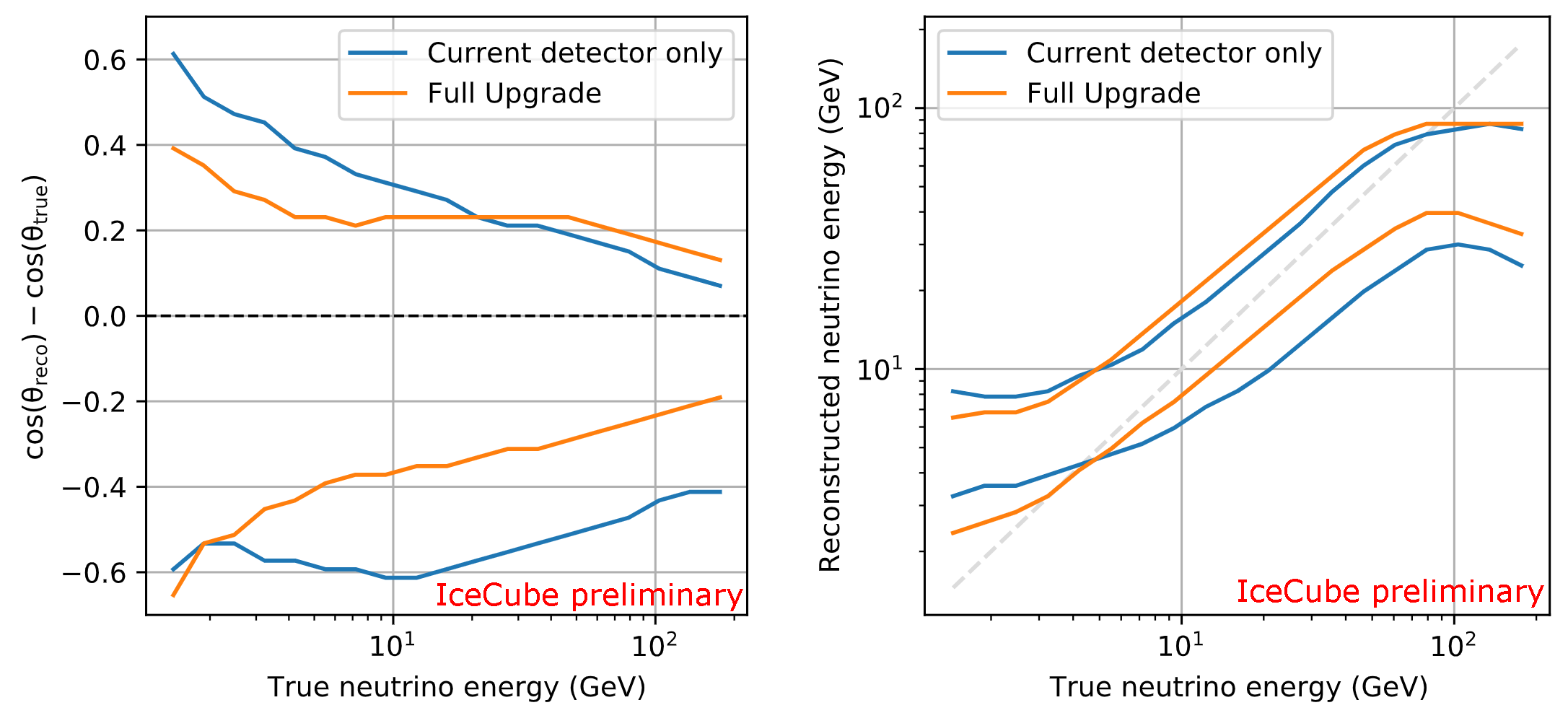}
    \caption{Bands comparing the resolution vs energy in the zenith angle $\cos(\theta)$ (left) and the neutrino energy $E$ (right) for the IceCube Upgrade ("Full Upgrade") and with the Upgrade strings removed ("Current detector only"). 68\% of the best reconstructed events lie between the respective bands. The dashed lines indicated the target values.}
    \label{fig:upgrade_res}
\end{figure}

We can see in the results in Fig.\ \ref{fig:upgrade_res} that there is an overall improvement in resolution, both in the width of the bands, as well as the centering around the target values. However, due the premature event selection, these results are not final and not representative of the expected resolution for the IceCube Upgrade.
Over all tested samples, $\mathrm{\sigma_{\cos(\theta),\ Upgrade}} = 0.35$ and $\mathrm{\sigma_{\cos(\theta),\ current}} = 0.40$. In the true neutrino energy range between 10 GeV and 100 GeV we receive $\sigma_{\mathrm{log_{10}(E),\ Upgrade}}= 0.22$ and $\sigma_{\mathrm{log_{10}(E),\ current}}= 0.29$. Additionally, in Fig.\ \ref{fig:upgrade_roc} the receiver-operator-characteristic curves of the track--cascade identifier are shown for both cases. We can see from their shapes, as well from the area under the curve values of $\mathrm{AUC_{Upgrade}} = 0.83$ and $\mathrm{AUC_{current}} = 0.75$, that there is also an improvement in the distinguishing power with the additional strings.


\section{Conclusion}

Graph Neural Networks offer a method for reconstructing and classifying neutrino events for both the IceCube and the IceCube Upgrade. With our approach we can encode the detector pulses without major preprocessing such as summary statistics or geometry transformations. Compared to our current baseline algorithms, our results not only outpace them in resolution, but also in speed. We will look to improve our performance and investigate the use of our networks for the event selection in the near future.

\bibliographystyle{ICRC}
\bibliography{references}



\clearpage
\section*{Full Author List: IceCube Collaboration}

\scriptsize
\noindent
R. Abbasi$^{17}$,
M. Ackermann$^{59}$,
J. Adams$^{18}$,
J. A. Aguilar$^{12}$,
M. Ahlers$^{22}$,
M. Ahrens$^{50}$,
C. Alispach$^{28}$,
A. A. Alves Jr.$^{31}$,
N. M. Amin$^{42}$,
R. An$^{14}$
% ...
\\

\noindent
% ...
$^{12}$ Universit{\'e} Libre de Bruxelles, Science Faculty CP230, B-1050 Brussels, Belgium \\
% ...
$^{14}$ Department of Physics and Laboratory for Particle Physics and Cosmology, Harvard University, Cambridge, MA 02138, USA \\
% ...
$^{17}$ Department of Physics, Loyola University Chicago, Chicago, IL 60660, USA \\
$^{18}$ Dept. of Physics and Astronomy, University of Canterbury, Private Bag 4800, Christchurch, New Zealand \\
% ...
$^{22}$ Niels Bohr Institute, University of Copenhagen, DK-2100 Copenhagen, Denmark \\
% ...
$^{28}$ D{\'e}partement de physique nucl{\'e}aire et corpusculaire, Universit{\'e} de Gen{\`e}ve, CH-1211 Gen{\`e}ve, Switzerland \\
% ...
$^{31}$ Karlsruhe Institute of Technology, Institute for Astroparticle Physics, D-76021 Karlsruhe, Germany  \\
% ...
$^{42}$ Bartol Research Institute and Dept. of Physics and Astronomy, University of Delaware, Newark, DE 19716, USA \\
% ...
$^{50}$ Oskar Klein Centre and Dept. of Physics, Stockholm University, SE-10691 Stockholm, Sweden \\
% ...
$^{59}$ DESY, D-15738 Zeuthen, Germany \\
% ...

\subsection*{Acknowledgements}

\noindent
USA {\textendash} U.S. National Science Foundation-Office of Polar Programs,
U.S. National Science Foundation-Physics Division,
U.S. National Science Foundation-EPSCoR,
Wisconsin Alumni Research Foundation,
Center for High Throughput Computing (CHTC) at the University of Wisconsin{\textendash}Madison,
Open Science Grid (OSG),
Extreme Science and Engineering Discovery Environment (XSEDE),
Frontera computing project at the Texas Advanced Computing Center,
U.S. Department of Energy-National Energy Research Scientific Computing Center,
Particle astrophysics research computing center at the University of Maryland,
Institute for Cyber-Enabled Research at Michigan State University,
and Astroparticle physics computational facility at Marquette University;
Belgium {\textendash} Funds for Scientific Research (FRS-FNRS and FWO),
FWO Odysseus and Big Science programmes,
and Belgian Federal Science Policy Office (Belspo);
Germany {\textendash} Bundesministerium f{\"u}r Bildung und Forschung (BMBF),
Deutsche Forschungsgemeinschaft (DFG),
Helmholtz Alliance for Astroparticle Physics (HAP),
Initiative and Networking Fund of the Helmholtz Association,
Deutsches Elektronen Synchrotron (DESY),
and High Performance Computing cluster of the RWTH Aachen;
Sweden {\textendash} Swedish Research Council,
Swedish Polar Research Secretariat,
Swedish National Infrastructure for Computing (SNIC),
and Knut and Alice Wallenberg Foundation;
Australia {\textendash} Australian Research Council;
Canada {\textendash} Natural Sciences and Engineering Research Council of Canada,
Calcul Qu{\'e}bec, Compute Ontario, Canada Foundation for Innovation, WestGrid, and Compute Canada;
Denmark {\textendash} Villum Fonden and Carlsberg Foundation;
New Zealand {\textendash} Marsden Fund;
Japan {\textendash} Japan Society for Promotion of Science (JSPS)
and Institute for Global Prominent Research (IGPR) of Chiba University;
Korea {\textendash} National Research Foundation of Korea (NRF);
Switzerland {\textendash} Swiss National Science Foundation (SNSF);
United Kingdom {\textendash} Department of Physics, University of Oxford.

\end{document}